\documentclass[aps,prb,amsmath,twocolumn]{revtex4}
\usepackage{bm}
\usepackage{graphicx}

\DeclareMathOperator{\sgn}{sgn}
\DeclareMathOperator{\Img}{\mathrm{Im}}
\DeclareMathOperator{\Rea}{\mathrm{Re}}

\begin{document}
\title{Josephson current and Andreev states in superconductor-half metal-superconductor heterostructures}
\author{Artem V. Galaktionov}
\affiliation{I.E. Tamm Department of Theoretical Physics, P.N.
Lebedev Physics Institute, 119991 Moscow, Russia}
\author{Mikhail S. Kalenkov}
\affiliation{I.E. Tamm Department of Theoretical Physics, P.N.
Lebedev Physics Institute, 119991 Moscow, Russia}
\author{Andrei D. Zaikin}
\affiliation{Forschungszentrum Karlsruhe, Institut f\"ur Nanotechnologie,
76021, Karlsruhe, Germany}
\affiliation{I.E. Tamm Department of Theoretical Physics, P.N.
Lebedev Physics Institute, 119991 Moscow, Russia}

\begin{abstract}
We develop a detailed microscopic theory describing dc Josephson
effect and Andreev bound states in superconducting junctions with
a half-metal. In such systems the supercurrent is caused by
triplet pairing states emerging due to spin-flip scattering at the
interfaces between superconducting electrodes and the half-metal.
For sufficiently clean metals we provide a detailed non-perturbative
description of the Josephson current at arbitrary transmissions
and spin-flip scattering parameters for both interfaces. Our
analysis demonstrates that the behavior of both the Josephson
current and Andreev bound states crucially depends on the strength
of spin-flip scattering showing a rich variety of features which
can be tested in future experiments.
\end{abstract}

\pacs{74.50.+r, 73.40.-c,  74.45.+c}
\maketitle

\section{Introduction}
\label{secintr}

Recent experiments \cite{Keizer} strongly indicate the possibility
to realize non-vanishing supercurrent across sufficiently thick
half-metal (H) layers embedded in-between two $s$-wave BCS
superconductors (S). This physical situation appears rather
non-trivial. Indeed, in conventional SNS junctions (N stands for
spin-isotropic normal metal) the supercurrent is carried by
(spin-singlet) Cooper pairs penetrating into the N-metal layer
from both superconductors due to the proximity effect. However,
half-metals are fully spin polarized materials acting as
insulators for electrons with one of the two spin directions.
Hence, penetration of spin-singlet electron pairs into half metals
should be prohibited and no supercurrent would be possible.

Recently it was realized \cite{Eschrig03} that this situation
changes qualitatively if one allows for spin-flip scattering at HS
interfaces. Such scattering enables conversion of spin-singlet
pairing in S-electrodes into spin-triplet pairing in a half-metal.
In this way superconducting correlations can survive even in a
half-metal ferromagnet, thus ''unblocking'' the supercurrent
across SHS junctions. Subsequent numerical and analytical studies
\cite{Sasha,Eschrig07,EL} confirmed this physical picture also
extending it to structures with disorder. In particular, it was
argued \cite{EL} that depending on the degree of disorder in the
H-metal the origin of triplet pairing there can change from the
$p$-wave type to the odd-frequency one \cite{BVE}. It was also
realized that the presence of triplet pairing in strong ferromagnets
and in half-metals can cause the so-called $\pi$-junction behavior of the
system \cite{VBE}.

Despite all these important developments the issue is yet far from
settled. The goal of this paper is to work out a complete theory
of dc Josephson effect in clean SHS heterostructures at arbitrary
transmissions of HS interfaces. Electron scattering at these
interfaces will be described by the most general scattering
matrices which fully account for spin-flip processes. Depending on
the system parameters we will find a rich variety of the results
both for the temperature dependence of the supercurrent and for
the current-phase relation in SHS junctions demonstrating crucial
importance of spin-flip scattering at HS interfaces. In addition
we will also address Andreev level quantization and show that this
phenomenon in SHS junctions acquires qualitatively new features
which are not present in conventional SNS structures.

The structure of the paper is as follows. In Sec. \ref{secquas} we
will employ the quasiclassical formalism which enables us to
exactly evaluate the Josephson current in SHS structures with many
conducting channels at any transmissions of SH interfaces. In Sec.
\ref{secbeyond} we will develop a more general approach which also
accounts for resonant effects and allows to include structures
with few conducting channels into consideration. Within this
approach we will describe both the Josephson current and Andreev
levels in SHS junctions and establish the correspondence to the
results derived in Sec. \ref{secquas}. In Sec. \ref{secsumm} we
will briefly summarize our main observations.

\section{Quasiclassical analysis}
\label{secquas}
In this section we will consider a general model of a clean SNS
junction with spin-active interfaces. We will then specify the
scattering matrices of NS interfaces in a way appropriate to
describe SHS heterostructures. Here we will treat the systems with
many conducting channels. For this reason it will be sufficient to
employ the quasiclassical formalism of energy-integrated Matsubara
Green functions \cite{Eil,BWBSZ}.

\begin{figure*}
\centerline{
\includegraphics{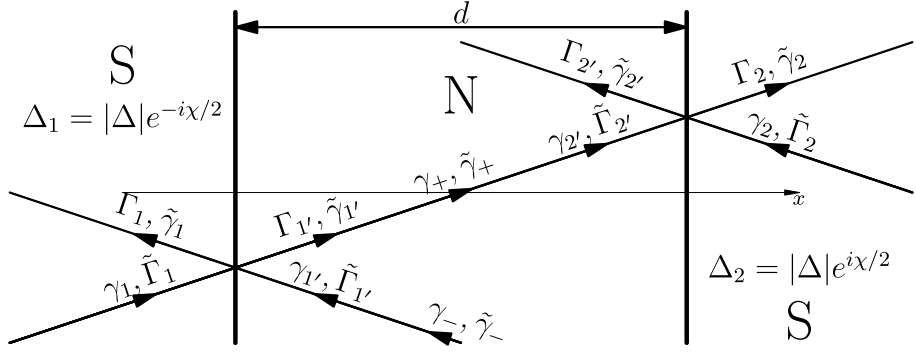}
} \caption{SNS junction and Riccati amplitudes in the clean limit.
The functions $\gamma_i$, $\Gamma_i$ $\tilde\gamma_i$
$\tilde\Gamma_i$ are Riccati amplitudes at the corresponding NS
interface. $\gamma_{\pm}$ and $\tilde\gamma_{\pm}$ are Riccati
amplitudes in the middle of the normal metal layer. Quasiparticle
momentum directions are indicated by arrows.} \label{traject}
\end{figure*}

\subsection{Riccati parameterization}
\label{subsecric}

In the ballistic limit the Eilenberger equations take the form
\begin{equation}
\left[
i\omega_n \hat\tau_3-
\hat\Delta(\bm{r}),
\hat g (\bm{p}_F, \omega_n, \bm{r})
\right]+
i\bm{v}_F \nabla \hat g (\bm{p}_F, \omega_n, \bm{r}) =0,
\label{Eil}
\end{equation}
where $[\hat a, \hat b]= \hat a\hat b - \hat b \hat
a$, $\omega_n = \pi T (2n+1)$ is Matsubara frequency, $\bm{p}_F=m\bm{v}_F$ is the
electron Fermi momentum vector and $\hat\tau_3$ is the Pauli matrix in Nambu
space. The function  $\hat g$ also obeys the normalization condition
$\hat g^2=1$. Green function $\hat g$ and $\hat\Delta$ are $4\times4$ matrices in
Nambu and spin spaces. In Nambu space they can be parameterized as
\begin{equation}
        \hat g =
        \begin{pmatrix}
                g & f \\
                \tilde f & \tilde g \\
        \end{pmatrix}, \quad
        \hat\Delta=
        \begin{pmatrix}
                0 & \Delta i\sigma_2 \\
                \Delta^* i\sigma_2& 0 \\
        \end{pmatrix},
\end{equation}
where $g$, $f$, $\tilde f$, $\tilde g$ are $2\times2$ matrices in
the spin space, $\Delta$ is the BCS order parameter and $\sigma_i$
are Pauli matrices. For simplicity we will only consider the case
of spin-singlet isotropic pairing in superconducting electrodes.
As usually, the superconducting order parameter in the normal layer
is set to be equal to zero.

The equilibrium current density is defined by the standard
relation
\begin{equation}
\bm{j}(\bm{r})= e N_0\pi T \sum\limits_{\omega_n >0}
\Img
\left< \bm{v}_F \mathrm{Sp} [\hat \tau_3 \hat g(\bm{p}_F,
\omega_n, \bm{r})] \right>,
\label{current}
\end{equation}
where $N_0=mp_F/2\pi^2$ is the density of states at the Fermi
level and angular brackets $\left< ... \right>$ denote averaging
over the Fermi momentum.

The above matrix Green functions can be conveniently parameterized
\cite{Eschrig00} by the two Riccati amplitudes $\gamma$ and
$\tilde \gamma$:
\begin{equation}
\hat g=
\begin{pmatrix}
(1-\gamma \tilde \gamma)^{-1}(1+\gamma \tilde \gamma) &
2(1-\gamma \tilde \gamma)^{-1}\gamma \\
-2(1-\tilde \gamma  \gamma)^{-1} \tilde \gamma
& -(1-\tilde \gamma  \gamma)^{-1}(1+ \tilde \gamma \gamma) \\
\end{pmatrix}.
\label{gparam}
\end{equation}
With the aid of the above parameterization one can identically
transform the quasiclassical equations \eqref{Eil} into the
following set of decoupled equations for Riccati amplitudes
\cite{Eschrig00}
\begin{gather}
i\bm{v}_F \nabla \gamma + 2i\omega \gamma = \gamma \Delta^* i\sigma_2\gamma-\Delta i\sigma_2,
\label{eqgamma}
\\
i\bm{v}_F \nabla \tilde\gamma - 2i\omega\tilde\gamma =
\tilde\gamma\Delta i\sigma_2\tilde\gamma -\Delta^* i\sigma_2.
\label{eqtildegamma}
\end{gather}

Solving Eqs. \eqref{eqgamma} and \eqref{eqtildegamma} inside the normal metal
we obtain following relations between Riccati amplitudes
\begin{gather}
\gamma_+=\Gamma_{1'}\exp(-\omega_n d/|v_{Fx}|)=\gamma_{2'}\exp(\omega_n d/|v_{Fx}|), \\
\gamma_-=\gamma_{1'}\exp(\omega_n d/|v_{Fx}|)=\Gamma_{2'}\exp(-\omega_n d/|v_{Fx}|),\\
\tilde\gamma_+=\tilde\gamma_{1'}\exp(\omega_n d/|v_{Fx}|)=\tilde\Gamma_{2'}\exp(-\omega_n d/|v_{Fx}|), \\
\tilde\gamma_-=\tilde\Gamma_{1'}\exp(-\omega_n d/|v_{Fx}|)=\tilde\gamma_{2'}\exp(\omega_n d/|v_{Fx}|),
\end{gather}
where $\gamma_{i'}$, $\Gamma_{i'}$, $\tilde\gamma_{i'}$,
$\tilde\Gamma_{i'}$ are Riccati amplitudes at the corresponding
interface and $\gamma_{\pm}$ and $\tilde\gamma_{\pm}$ are Riccati
amplitudes in the middle of the normal metal slab (see Fig.
\ref{traject} for details), $d$ is a distance between two NS
interfaces.

Deep inside the superconducting electrodes we apply the following 
asymptotic conditions
\begin{gather}
\gamma_1 = - \sigma_2 a(\omega)e^{-i\chi/2}, \quad
\tilde\gamma_1 =  \sigma_2 a(\omega)e^{i\chi/2},\\
\gamma_2 = - \sigma_2 a(\omega)e^{i\chi/2}, \quad
\tilde\gamma_2 =  \sigma_2 a(\omega)e^{-i\chi/2},
\label{asy}
\end{gather}
where $a(\omega) = ( \sqrt{ \omega^2 + |\Delta|^2 } - \omega )/ |\Delta|$ and
$\chi$ is the superconducting phase difference across the junction. Here and
below we assume that the order parameter is spatially uniform inside
superconducting electrodes. This choice can always be parametrically justified 
by assuming the proper junction geometry and/or interface transmission
values. For simplicity we also assume that the absolute values of the 
superconducting order parameter are identical in both S-electrodes. 
Generalization of our approach to the case of anisotropic pairing and 
asymmetric electrodes with $|\Delta_1|\neq|\Delta_2|$ is straightforward.

\subsection{Boundary conditions at metallic interfaces}
\label{subsecbound}

The above quasiclassical equations should be supplemented by
appropriate boundary conditions at the interfaces. In the case of
specularly reflecting spin-degenerate interfaces these conditions
were derived by Zaitsev \cite{Zaitsev} and later generalized to
spin-active interfaces in Ref. \onlinecite{Millis88}.

Similarly to Ref. \onlinecite{KZ07} it will be convenient for us to use the boundary conditions
formulated directly in terms of Riccati amplitudes. Let us
consider the first NS interface and explicitly specify the
relations between Riccati amplitudes for incoming and outgoing
electron trajectories, see Fig.~\ref{traject}. For instance, the
boundary conditions for $\Gamma_1$, can be written in the form
\cite{Fogelstrom00,Zhao04}
\begin{gather}
\Gamma_1= r_{1l}\gamma_1 \underline{S}_{11}^+ +
t_{1l}\gamma_{1'} {\underline{S}}_{11'}^+,
\\
\tilde\Gamma_1= \underline{S}_{11}^+ \tilde\gamma_1 \tilde r_{1r}+
\underline{S}_{1'1}^+ \tilde\gamma_{1'} \tilde t_{1r}.
\end{gather}
Here we defined the transmission ($t$) and reflection  ($r$) amplitudes as:
\begin{gather}
r_{1l}=[\beta_{1'1}^{-1}S_{11}^+ - \beta_{1'1'}^{-1}S_{11'}^+]^{-1}\beta_{1'1}^{-1},
\\
\tilde r_{1r}=\beta_{11'}^{-1}[S_{11}^+ \beta_{11'}^{-1} - S_{1'1}^+\beta_{1'1'}^{-1}]^{-1},
\\
t_{1l}=-[\beta_{1'1}^{-1}S_{11}^+ - \beta_{1'1'}^{-1}S_{11'}^+]^{-1}\beta_{1'1'}^{-1},
\\
\tilde t_{1r}=-\beta_{1'1'}^{-1}[S_{11}^+ \beta_{11'}^{-1} - S_{1'1}^+\beta_{1'1'}^{-1}]^{-1},
\end{gather}
where
\begin{gather}
\beta_{ij}=S_{ij}^+ - \gamma_j \underline{S}_{ij}^+ \tilde\gamma_i.
\end{gather}
Matrices $S_{ij}$ and $\underline{S}_{ij}$ are building blocks of the full electron and hole interface S-matrices\cite{Millis88}
\begin{equation}
\mathcal{S}=
\begin{pmatrix}
S_{11} & S_{11'} \\
S_{1'1} & S_{1'1'} \\
\end{pmatrix}, \quad
\underline{\mathcal{S}}=
\begin{pmatrix}
\underline{S}_{11} & \underline{S}_{11'} \\
\underline{S}_{1'1} & \underline{S}_{1'1'} \\
\end{pmatrix}
\end{equation}

Boundary conditions for $\Gamma_{1'}$, $\tilde \Gamma_{1'}$ can be
obtained from the above equations simply by replacing $1
\leftrightarrow 1'$. Boundary conditions describing electron
scattering at the second interface are formulated analogously.

Combining the above relations between Riccati amplitudes we obtain
matrix quadratic equations for the matrices $\eta_{\pm} = -i
\gamma_{\pm} \sigma_2$ and $\tilde\eta_{\pm} = i\sigma_2\tilde\gamma_{\pm} $
\begin{multline}
\eta_+ \Bigl\{ c_2 a_1 \exp(-\omega_n d/|v_{Fx}|) + a_2 b_1\exp(\omega_n d/|v_{Fx}|) \Bigr\}\eta_+
-\\-
\eta_+ \Bigl\{c_2 c_1 \exp(-2\omega_n d/|v_{Fx}|) + a_2 d_1 \Bigr\}
+\\+
\Bigl\{ d_2 a_1 + b_2 b_1 \exp(2\omega_n d/|v_{Fx}|) \Bigr\}\eta_+
-\\-
\Bigl\{ d_2 c_1 \exp(-\omega_n d/|v_{Fx}|) + b_2 d_1 \exp(\omega_n d/|v_{Fx}|) \Bigr\}=0,
\label{quadeq}
\end{multline}
\begin{multline}
\tilde\eta_+ \Bigl\{ d_2 c_1 \exp(-\omega_n d/|v_{Fx}|) + b_2 d_1 \exp(\omega_n d/|v_{Fx}|) \Bigr\}\tilde\eta_+
-\\-
\tilde\eta_+\Bigl\{ d_2 a_1 + b_2 b_1 \exp(2\omega_n d/|v_{Fx}|) \Bigr\}
+\\+
\Bigl\{c_2 c_1 \exp(-2\omega_n d/|v_{Fx}|) + a_2 d_1 \Bigr\}\tilde\eta_+
-\\-
\Bigl\{ c_2 a_1 \exp(-\omega_n d/|v_{Fx}|) + a_2 b_1\exp(\omega_n d/|v_{Fx}|) \Bigr\}
=0,
\label{quadeq1}
\end{multline}
where we introduced the following $2\times 2$ matrices
\begin{gather}
a_1=\left[i\sigma_2\underline{S}_{11'}^+ \tilde\gamma_1\beta_{11}^{-1} S_{1'1}^+ \right],
\label{defa1}
\\
b_1=\left[S_{1'1'}^+ -S_{11'}^+\beta_{11}^{-1} S_{1'1}^+\right],
\label{defb1}
\\
c_1=\left[\sigma_2(\underline{S}_{1'1'}^+ +\underline{S}_{11'}^+ \tilde\gamma_1 \beta_{11}^{-1}
\gamma_1 \underline{S}_{1'1}^+ ) \sigma_2\right],
\label{defc1}
\\
d_1=\left[S_{11'}^+\beta_{11}^{-1} \gamma_1 \underline{S}_{1'1}^+ i\sigma_2\right],
\label{defd1}
\\
a_2=\left[i\sigma_2\underline{S}_{22'}^+ \tilde\gamma_2\beta_{22}^{-1} S_{2'2}^+ \right],
\label{defa2}
\\
b_2=\left[S_{2'2'}^+ -S_{22'}^+\beta_{22}^{-1} S_{2'2}^+\right],
\label{defb2}
\\
c_2=\left[\sigma_2 ( \underline{S}_{2'2'}^+ +\underline{S}_{22'}^+ \tilde\gamma_2 \beta_{22}^{-1}
\gamma_2 \underline{S}_{2'2}^+ )\sigma_2\right],
\label{defc2}
\\
d_2=\left[S_{22'}^+\beta_{22}^{-1} \gamma_2 \underline{S}_{2'2}^+ i\sigma_2 \right].
\label{defd2}
\end{gather}

Matrix equations for the $\eta_-$ and $\tilde\eta_-$ can be
obtained from Eqs. \eqref{quadeq} and \eqref{quadeq1} by
substituting $\eta_+ \rightarrow \eta_-$, $\tilde\eta_+
\rightarrow \tilde\eta_-$ and interchanging of indices
$1\leftrightarrow2$.

For arbitrary interface $S$-matrices the matrix equations
\eqref{quadeq} and \eqref{quadeq1} can be reduced to scalar quartic
equations with very cumbersome general solutions. These solutions,
however, become simpler for some particular interface models. Here
we will stick to the case of SHS junctions in which we should
specify the scattering $S$-matrix for the interface between
spin-isotropic normal metal (or superconductor) and fully spin
polarized ferromagnet. We will demonstrate that in this case Eqs.
\eqref{quadeq} and \eqref{quadeq1} can be solved in a transparent
and compact way.

\subsection{Scattering matrices}
\label{subsecscat}

Let specify the scattering $S$-matrices for both SH interfaces.
For simplicity we will assume that these matrices depend only on
the incidence angle but not on the azimuthal one. Then the
following relation between electron and hole $S$-matrices holds:
$\mathcal{S}=\underline{\mathcal{S}}^T$. For a half-metal slab between two
superconducting electrodes the corresponding interface scattering
matrices contain $3\times 3$ nontrivial sub-matrices, i.e.
\begin{equation}
\mathcal{S}=
\begin{pmatrix}
\cdots & \cdots & \cdots & 0\\
\cdots & \cdots & \cdots & 0\\
\cdots & \cdots & \cdots & 0\\
0 & 0 & 0 & 1 \\
\end{pmatrix}.
\label{shalf}
\end{equation}

It is straightforward to demonstrate that the Josephson current is
invariant under the following transformation of the $S$-matrices:
\begin{gather}
\mathcal{S}_1\rightarrow
\begin{pmatrix}
U_1 & 0 \\
0 & V \\
\end{pmatrix}
\mathcal{S}_1
\begin{pmatrix}
U_1^+ & 0 \\
0 & V^+ \\
\end{pmatrix},
\label{trans1}
\\
\underline{\mathcal{S}}_1\rightarrow
\begin{pmatrix}
U_1^* & 0 \\
0 & V^* \\
\end{pmatrix}
\underline{\mathcal{S}}_1
\begin{pmatrix}
U_1^T & 0 \\
0 & V^T \\
\end{pmatrix},
\label{trans2}
\\
\mathcal{S}_2\rightarrow
\begin{pmatrix}
U_2 & 0 \\
0 & V \\
\end{pmatrix}
\mathcal{S}_2
\begin{pmatrix}
U_2^+ & 0 \\
0 & V^+ \\
\end{pmatrix},
\label{trans3}
\\
\underline{\mathcal{S}}_2\rightarrow
\begin{pmatrix}
U_2^* & 0 \\
0 & V^* \\
\end{pmatrix}
\underline{\mathcal{S}}_2
\begin{pmatrix}
U_2^T & 0 \\
0 & V^T \\
\end{pmatrix},
\label{trans4}
\end{gather}
where $U_1,U_2,V\in SU(2)$. With the aid of the above
transformation we can always reduce the $S$-matrices to the
following form
\begin{equation}
\mathcal{S}=
\begin{pmatrix}
\alpha e^{i\zeta} \cos\nu  & \beta e^{i\zeta} & \alpha e^{-2i\zeta} \sin\nu  & 0\\
-\beta^*e^{i\zeta} \cos\nu & \alpha^* e^{i\zeta} & -\beta^* e^{-2i\zeta} \sin\nu  & 0\\
-e^{i\zeta} \sin\nu & 0  & e^{-2i\zeta} \cos\nu  & 0\\
0 & 0 & 0& 1\\
\end{pmatrix},
\label{shalf2}
\end{equation}
where $|\alpha|^2+|\beta|^2=1$ and $\nu$, $\zeta$ are real. We also note
that in general the first and the second interfaces are characterized
by two different sets of parameters $\alpha$, $\beta$, $\nu$ and $\zeta$,
i.e. these interfaces are described by different scattering matrices. 

Here and below the parameter
$\sin^2\nu$ defines an effective interface transmission. The
normal state differential conductance ($dI/dV$) for the metallic
interface described by the $\mathcal{S}$-matrix \eqref{shalf2} is
proportional to $\sin^2\nu$
\begin{equation}
G_{NN}=\dfrac{\mathcal{A}e^2}{2\pi} \int\limits_{|p_\parallel|<p_F} \frac{d^2
p_\parallel}{(2\pi)^2} \sin^2 \nu = \dfrac{e^2}{2\pi}\sum_k \sin^2 \nu
\end{equation}
where the index $k$ labels conducting channels of our junction. The
limit $\nu=0$ corresponds to completely impenetrable interfaces. The parameter
$\beta$ is responsible for spin-flip scattering of electrons at the interface.

Finally, we point out that by virtue of Eqs. \eqref{trans1}-\eqref{trans4} 
the scattering matrix employed 
in the analysis of Ref. \onlinecite{Eschrig07} can be reduced to
our expression \eqref{shalf2} provided we identify
\begin{gather}
\cos\nu=1-\dfrac{t_{\uparrow\uparrow}^2+t_{\downarrow\uparrow}^2}{2W}, \quad
\sin\nu=\dfrac{\sqrt{t_{\uparrow\uparrow}^2+t_{\downarrow\uparrow}^2}}{W}, \\
\alpha=-\dfrac{t_{\uparrow\uparrow}^2}{t_{\uparrow\uparrow}^2+t_{\downarrow\uparrow}^2}
e^{i\theta/2} -\dfrac{t_{\downarrow\uparrow}^2}{t_{\downarrow\uparrow}^2+t_{\downarrow\uparrow}^2}
e^{-i\theta/2}, \\
\beta=2 i \dfrac{t_{\uparrow\uparrow}t_{\downarrow\uparrow}
}{t_{\uparrow\uparrow}^2+t_{\downarrow\uparrow}^2}\sin (\theta/2)
e^{-i(\theta_{\uparrow\uparrow}+\theta_{\downarrow\uparrow})},\quad
\zeta=0,
\end{gather}
where we use the notations from Ref. \onlinecite{Eschrig07}. 

\subsection{Josephson current}
\label{subsecjos}

Now we are ready to evaluate the Josephson current. In SNS systems
with interface $S$-matrices of the form \eqref{shalf2} the
matrices $a_i,b_i,c_i,d_i,\eta_{\pm},\tilde\eta_{\pm}$ have
the following structure
\begin{gather}
a_i=
\begin{pmatrix}
0 & 0 \\
a_i & 0 \\
\end{pmatrix}, \quad
b_i=
\begin{pmatrix}
b_i & 0 \\
0 & 1 \\
\end{pmatrix}, \quad
c_i=
\begin{pmatrix}
1 & 0 \\
0 & c_i \\
\end{pmatrix},
\\
d_i=
\begin{pmatrix}
0 & d_i \\
0 & 0 \\
\end{pmatrix}, \quad
\eta_{\pm}=\begin{pmatrix}
0 & \eta_{\pm} \\
0 & 0 \\
\end{pmatrix}, \quad
\tilde\eta_{\pm}=\begin{pmatrix}
0 & 0 \\
\tilde\eta_{\pm} & 0 \\
\end{pmatrix}.
\end{gather}
Since all the matrices $a_i,b_i,c_i,d_i,\eta_{\pm},\tilde\eta_{\pm}$
have only one nontrivial matrix element it suffices to denote this
element by the same symbol as the corresponding matrix
itself. Then from Eqs. \eqref{quadeq}, \eqref{quadeq1}  we derive a simple quadratic
equations for the scalar variables $\eta_{\pm}$ and $\tilde\eta_{\pm}$. Resolving this equation
and making use of the Riccati parameterization described in Sec.
\ref{subsecric} and \ref{subsecbound} we construct the Green-Eilenberger function for our
junction. Substituting this function into Eq. (\ref{current}) we
arrive at the following general result for the Josephson current
\begin{equation}
I(\chi)= -4e \mathcal{A} T \sum\limits_{\omega_n>0}
\int\limits_{|p_\parallel|<p_F} \frac{d^2 p_\parallel}{(2\pi)^2}
\dfrac{a^2 (1-a^2)^2 {\cal D}_{12}\sin\tilde\chi}{
 Q(\omega_n)}
\label{jshs}
\end{equation}
where ${\cal D}_{12}=|\beta_1| |\beta_2| \sin^2\nu_1 \sin^2\nu_2$,
${\cal A}$ is the junction cross section,
\begin{widetext}
\begin{multline}
 Q(\omega)=
\Biggl\{ \Biggl[ 2a^2 (1-a^2)^2 |\beta_1| |\beta_2| \sin^2\nu_1
\sin^2\nu_2 \cos\tilde\chi -\\- \left((1 - a^2)(1 - a^2
\cos^2\nu_1) + a^2 \left|\tilde\alpha_1\right|^2\right) \left((1 -
a^2)(1 - a^2 \cos^2\nu_2) +a^2
\left|\tilde\alpha_2\right|^2\right) \exp(2\omega_n d/|v_{Fx}|)
-\\- \left((1-a^2)(-a^2 + \cos^2\nu_1) +a^2
\left|\tilde\alpha_1\right|^2\right) \left((1-a^2)(-a^2 +
\cos^2\nu_2) +a^2 \left|\tilde\alpha_2\right|^2\right)
\exp(-2\omega_n d/|v_{Fx}|)\Biggr]^2 -\\- 4 \left|(1-a^2)^2
\cos\nu_1 + a^2 \tilde\alpha_1^2 \right|^2 \left|(1-a^2)^2
\cos\nu_2 + a^2 \tilde\alpha_2^2 \right|^2 \Biggr\}^{1/2},
\label{Q}
\end{multline}
\end{widetext}
and
\begin{equation}
\tilde\chi=\chi+3\zeta_2 + \arg\beta_2 -3\zeta_1 -\arg\beta_1,
\quad \tilde\alpha_i=\alpha_i +\alpha_i^*\cos\nu_i.
\label{betazeta}
\end{equation}
We observe that the Josephson current is proportional to the
parameter ${\cal D}_{12}$ which contains effective transmissions
of both interfaces $\sin^2\nu_1 \sin^2\nu_2$ and the two spin-flip
factors $|\beta_1|$ and $|\beta_2|$. The scattering matrix
parameters $\zeta_1$, $\arg\beta_1$, $\zeta_2$, and $\arg\beta_2$
enter into our result only in combination with the phase
difference $\chi$ \eqref{betazeta}.

Let us briefly analyze the above general expression for the
Josephson current. At small transmissions $\nu_1,\nu_2 \ll 1$ and
for sufficiently long junctions $d \gg \xi_0 (\nu_1^2+\nu_2^2)$
Eq.~\eqref{jshs} reduces to
\begin{widetext}
\begin{equation}
I(\chi)= -\dfrac{e \mathcal{A} T}{2} \sum_{\omega_n>0}
\int\limits_{|p_\parallel|<p_F} \frac{d^2 p_\parallel}{(2\pi)^2}
\dfrac{\Delta^2 \omega_n^2 |\beta_1| |\beta_2| \nu_1^2 \nu_2^2
\sin\tilde\chi}{
[\omega_n^2+\Delta^2(\Rea\alpha_1)^2][\omega_n^2+\Delta^2(\Rea\alpha_2)^2]}
\dfrac{1}{\sinh(2\omega_n d/|v_{Fx}|)},
\label{Eschrig}
\end{equation}
\end{widetext}
which matches with the analogous result derived in Ref.~\onlinecite{Eschrig07} 
provided we set $\Rea\alpha_{1,2} = 1$. For
$v_F/d \ll T \ll \Delta$ the integral in Eq. \eqref{jshs} is
dominated by the contribution of momenta values sufficiently close
to $p_\parallel=0$ and the dependence for the Josephson current on
the junction thickness $d$ acquires the standard exponential form
\begin{equation}
I= - \dfrac{4 e \mathcal{A} v_F p_F^2 T^2}{\pi \Delta^2 d}
\exp(-2\pi T d/v_F) \left. \dfrac{\mathcal{D}_{12}
\sin\tilde\chi}{\left|\tilde\alpha_1\right|^2
\left|\tilde\alpha_2\right|^2} \right|_{p_\parallel=0}.
\end{equation}
At lower temperatures $T \ll v_F/ d \ll \Delta$ Eq. \eqref{jshs}
yields the power-law dependence on $d$:
\begin{equation}
I= -\dfrac{7\zeta(3) e \mathcal{A} }{4\pi d^3\Delta^2}
\int\limits_{|p_\parallel|<p_F} \frac{d^2 p_\parallel}{(2\pi)^2}
 \dfrac{|v_{Fx}|^3{\cal D}_{12} \sin\tilde\chi}{
\left|\tilde\alpha_1\right|^2 \left|\tilde\alpha_2\right|^2 },
\label{d3}
\end{equation}
i.e. $I \propto 1/d^3$ at $T \to 0$. In the limit of the small
transparencies $\nu_{1,2} \ll 1$ or spin-flip factors $|\beta_1|,
|\beta_2| \ll 1$ the term in $Q(\omega)$ proportional to
$\cos\tilde\chi$ is irrelevant, and the current-phase relation
becomes purely sinusoidal
\begin{equation}
I(\chi)=-I_c \sin(\chi-\chi_0),
\end{equation}
where the phase shift $\chi_0$ depends on the scattering matrix
parameters (\ref{betazeta}) and, hence, in general can take any
value. Provided these values change randomly for different
conducting channels the net Josephson current across the system
can be significantly reduced. For symmetric interfaces the phase
shift $\chi_0$ is identically zero and the $\pi$-junction behavior
is realized.

Note that the expression (\ref{Eschrig}) formally diverges in the
limit of small $d$ illustrating insufficiency of the perturbative
(in the transmission) approach for the case of sufficiently short
SHS junctions. This divergence is, however, regularized within
the non-perturbative approach adopted here.
From Eqs. (\ref{jshs}), (\ref{Q}) we observe that for very short
junctions $d\ll \xi_0(\nu_1^2+\nu_2^2)$ (and for
$\Rea\alpha_1=\Rea\alpha_2$) the Josephson current scales with the
tunnel interface transmissions as
\begin{equation}
I \propto \frac{\nu_1^2\nu_2^2}{\nu_1^2+\nu_2^2}.
\end{equation}
Obviously this dependence cannot be derived within a simple
perturbative approach in $\nu_{1,2}$.
\begin{figure}
\centerline{
\includegraphics[width=75mm]{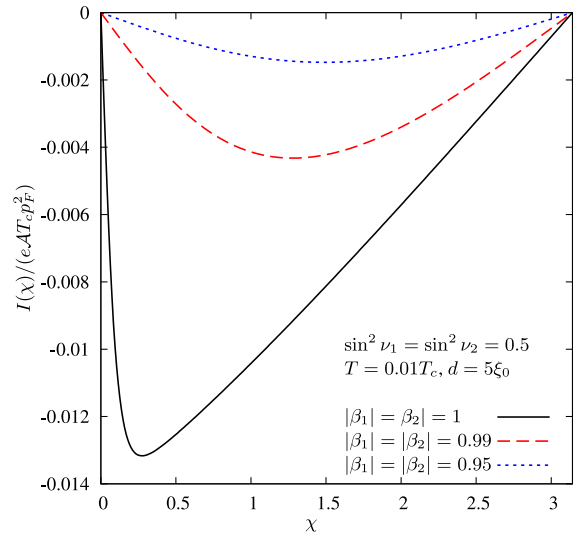}
}\caption{Phase dependence of the Josephson current at low
temperatures and different spin-flip factors $\beta$. For
simplicity we consider the case of identical interfaces and assume
that scattering parameters $\nu, \beta, \alpha, \zeta$ are
momentum independent. Parameter $\alpha$ is chosen to be real.
 The phase coherence length $\xi_0$ is
defined as $\xi_0=v_F/(2\pi T_c)$.} \label{j-chi}
\end{figure}

\begin{figure}
\centerline{
\includegraphics[width=75mm]{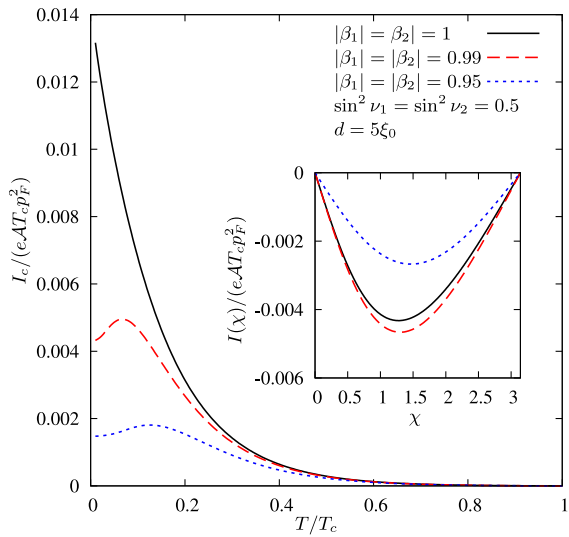}
}\caption{Critical Josephson current $I_c$ as a function of
temperature for different spin-flip factors $\beta$. The inset
shows the current-phase relation for $\beta_1=\beta_2=0.99$ and at
temperatures $T/T_c$=0.01 (solid line), 0.1 (dashed line), 0.2
(dotted line). For simplicity we consider the case of identical
interfaces and assume that scattering parameters $\nu, \beta,
\alpha, \zeta$ are momentum independent. Parameter $\alpha$ is
chosen to be real.} \label{j-T}
\end{figure}

A detailed analysis of analytical expressions for the current in
the limit of high interface transmissions will be postponed to the
next section. Here we only present Figs. \ref{j-chi} and \ref{j-T}
illustrating some key features of our general results
(\ref{jshs})-(\ref{betazeta}). The current-phase relation for
sufficiently thick SHS junction is depicted in Fig. \ref{j-chi}
at $T=0.01T_c$ for the case of highly transmitting identical
interfaces. We observe that within the interval of Josephson
phases $\chi$ ranging from zero to $\pi$ the current is always
negative. Strong deviations from the sinusoidal current-phase
relation emerge only provided the spin-flip parameter $|\beta|$ is
very close to unity which corresponds to (almost) complete
spin-flip scattering at both interfaces. Temperature dependence of
the critical Josephson current is shown in Fig. \ref{j-T} for
different values of the spin-flip parameter $|\beta|$. Similarly
to Refs. \onlinecite{Eschrig03,Eschrig07} in a wide parameter
range we find non-monotonous dependence of the critical current on
$T$ with a maximum typically below $0.2T_c$. We also observe that
this feature disappears as the spin-flip factor $|\beta|$
approaches unity, i.e. monotonous increase of $I_c$ with
decreasing temperature is expected in the limit of complete
spin-flip scattering at the HS interfaces.

\section{Going beyond quasiclassics}
\label{secbeyond}

The quasiclassical approach used so far provides an exact solution for the
problem in the limit of large number of conducting channels in our structure.
In this case the quantum mechanical phases, relevant, e.g., for resonance
effects, average out \cite{GZ}. Such averaging is justified in a number of
important physical situations, for instance, provided surface roughness of
metallic interfaces exceeds the Fermi wavelength. On the other hand, the many
channel limit is not the only one of physical relevance for the systems in
question. Modern experimental techniques enable one to study the Josephson
current through objects with few conducting channels with controllable change
of the scattering phase \cite{Nano}. This renders a motivation for calculating
the Josephson current through such objects. In such cases the above
quasiclassical formalism is in general insufficient, and more accurate
treatment becomes necessary \cite{GZ}. In addition, correct description of
Andreev states in SHS junctions also requires going beyond the quasiclassical
approach employed in Sec. \ref{secquas}.

\subsection{General formalism}
\label{subsecgorkov}

Below we will make use of the more general microscopic formalism
based of the Gorkov equations. Let us introduce the matrix
Matsubara Green functions \cite{AGD}
\begin{equation}
G_{ll'}(x_1,x_2,\tau_1-\tau_2)=-\left\langle{\rm T}_\tau
\psi_{l}(x_1,\tau_1)\overline{\psi}_{l'}(x_2,\tau_2) \right\rangle,
\end{equation}
where the indices are numbered as
\begin{equation}
\psi_1=\psi_{\uparrow},\quad \psi_2=\psi_{\downarrow},\quad
\psi_3=\overline{\psi}_{\uparrow}, \quad \psi_4=\overline{\psi}_{\downarrow}.
\end{equation}
As before, assuming singlet pairing in superconducting electrodes,
one can write down the standard Gorkov equations
\begin{equation}
\left(i\omega_n-\left[\check \epsilon(x_1)+\check
\Delta(x_1)\right]\right)\check
G(x_1,x_2,\omega_n)=\delta(x_1-x_2),
\label{fc}
\end{equation}
where we performed the Fourier transformation with respect to
$\tau_1-\tau_2$ introducing the Matsubara frequencies $\omega_n$.
The matrix operators $\check \epsilon,\check \Delta$ have the
structure
\begin{equation}
\check \epsilon=\left( \begin{array}{cccc} \epsilon & 0 & 0 & 0\\ 0 & \epsilon
& 0 &0\\ 0 & 0& -\epsilon& 0\\ 0 & 0& 0& -\epsilon \end{array}\right),\quad
\check \Delta=\left( \begin{array}{cccc} 0 & 0 & 0 & \Delta\\ 0 & 0 & -\Delta
&0\\ 0 & -\Delta^*& 0 & 0\\ \Delta^* & 0& 0& 0 \end{array}\right),
\end{equation}
where
\begin{equation}
\epsilon(x_1)=-\frac{1}{2m}\nabla_{x_1}^2-\frac{p_{Fx}^2}{2m}+V(x_1).
\end{equation}
Here $p_{Fx}$($v_{Fx}$) is the $x$-component of the Fermi momentum
(velocity) for a transmission channel and $V(x)$ stands for the
potential energy.

Inside the half-metal it is necessary to account for triplet pairing
which amounts to solving the two equations
\begin{gather}
(i\omega_n-\epsilon(x_1))G_{11}(x_1,x_2,\omega_n)=\delta(x_1-x_2),
\label{hme}\\ 
(i\omega_n+\epsilon(x_1))G_{31}(x_1,x_2,\omega_n)=0.
\end{gather}
The current flowing through the half-metal layer is given by the expression
\begin{equation}
I=\frac{ie}{2m}T\sum_{\omega_n,k}\left(
\nabla_{x_2}-\nabla_{x_1}\right)_{x_2\rightarrow x_1} G_{11}(x_1,x_2,\omega_n).
\end{equation}
Here and below an additional sum over the channel index $k$ implies summation
over all conducting channels of the junction. In the many channel limit this
summation can be reduced to the integral over the momentum
\begin{equation}
\sum_k\rightarrow  \mathcal{A}  \int\limits_{|p_\parallel|<p_F} \frac{d^2
p_\parallel}{(2\pi)^2},
\end{equation}
which we already encountered in Sec. \ref{secquas}.

 In what
follows we will make use of the standard approximation
\begin{equation}
\nabla^2_x[f(x)e^{\pm ip_{Fx}x}]
=
\left[ -p_{Fx}^2f(x) \pm 2ip_{Fx}
\partial_{x}f(x)
\right]
e^{\pm ip_{Fx}x},
\end{equation}
which is justified for any function $f(x)$ that varies smoothly on
atomic distances. We further exactly follow the derivation
\cite{GZ}. Let us fix the argument of the Green functions $x_2$
inside the half-metal layer and analyze their dependence on $x_1$.
For instance, we observe that the solution of Eq. (\ref{fc})
decaying deep inside the left superconductor has the form
\begin{multline}
\left(\begin{array}{c} G_{11}\\ G_{41}\end{array}\right)= \left(
\begin{array}{c} 1\\ -i e^{i\chi/2}a \end{array} \right)
e^{-ip_{Fx}x_{1}}e^{\kappa(x_1+(d/2))} y_1(x_2)
+\\+ 
\left(
\begin{array}{c} 1\\ ie^{i\chi/2} a^{-1}\end{array} \right)
e^{ip_{Fx}x_{1}}e^{\kappa(x_1+(d/2))} y_9(x_2).
\end{multline}
Here $y_1(x_2), y_9(x_2)$ are arbitrary functions, $a$ is defined
below Eq. (\ref{asy}) and $\kappa=\sqrt{\omega_n^2+
|\Delta|^2}/v_{Fx}$. The spatial decay of the functions
$(G_{21},G_{31})$ is described analogously.

A particular solution of Eq. (\ref{hme}) in the half-metal at
$\omega_n>0$  reads
\begin{gather}
G_{11}=-\frac{i}{ v_{Fx}}\exp\left[\left(i p_{Fx}-\frac{\omega_n}{v_{Fx}}
\right)|x_1-x_2|\right],
\\
G_{31}=0.
\end{gather}
Further calculation amounts to writing down the general solution
in the half-metal layer and to matching it with the decaying
solution in the superconducting reservoirs. This matching is made
with the help of the scattering matrices, that relate outgoing and
incoming waves. It is necessary to use two triads: The first one
is composed of the functions $G_{11}, G_{21}$ in the
superconductor and the function $G_{11}$ in the half-metal while the
second one comprises the functions $G_{31}, G_{41}$ of the
superconductor and the function $G_{31}$ of the half-metal. The
second triad accounts for the hole-like excitations, hence it
should be described by the transposed scattering matrix.

By matching these two triads at the left and the right
interfaces we arrive at twelve linear equations for the variables
$y_1,\dots y_{12}$
\begin{gather}
\left(\begin{array}{c} y_1\\ y_2\\ qy_3 \end{array} \right)=\hat S_1
\left(\begin{array}{c} y_{9}\\ y_{10}\\ z_1+ q^{-1}y_6 \end{array}\right),
\label{ls1}\\ \left(\begin{array}{c} y_4\\ y_5\\ qy_6 \end{array} \right)=\hat
S_2 \left(\begin{array}{c} y_{11}\\ y_{12}\\ z_2+ q^{-1}y_3
\end{array}\right),
\\
\left(\begin{array}{c}i  e^{i\chi/2} a y_2\\ -i e^{i\chi/2} a
y_1\\ q^{-1} y_7 \end{array} \right)=\hat S_1^T \left(\begin{array}{c}
-ie^{i\chi/2}a^{-1} y_{10}\\ i  e^{i\chi/2}a^{-1} y_{9}\\ q y_8
\end{array}\right), 
\\
\left(\begin{array}{c}i  e^{-i\chi/2} a
y_5\\ -i e^{-i\chi/2} a y_4\\ q^{-1} y_8 \end{array} \right)=\hat S_2^T
\left(\begin{array}{c} -i e^{-i\chi/2} a^{-1} y_{12}\\ i e^{-i\chi/2}a^{-1}
y_{11}\\ q y_7
\end{array}\right). \label{ls4}
\end{gather}
Here we keep $\omega_n>0$, denote $q=\exp\left(\omega_n d/2
v_{Fx}\right)$ and define the scattering matrices $\hat S_{1,2}$
as non-trivial $3\times 3$ sub-matrices in Eq. (\ref{shalf}).

The solution of Eqs. (\ref{ls1})-(\ref{ls4}) takes the form
\begin{equation}
y_3=U_1 z_1+U_2 z_2,\quad y_6=V_1 z_1+V_2 z_2.
\label{62}
\end{equation}
Then the contribution to the Josephson current defined by the
Green functions with positive Matsubara frequencies reads
\begin{equation}
I_+=ie T\sum_{\omega_n>0, k}q^{-1}\left( V_1- U_2\right).
\label{63}
\end{equation}
The contribution to the current from negative Matsubara frequencies $I_-$ is
determined analogously. It is straightforward to observe that the total
Josephson current acquires the form 
\begin{equation}
 I=I_++I_-=2{\rm Re}\,I_+, 
\end{equation}
i.e. it will be sufficient for our purposes to evaluate only the term $I_+$.

The matrices $S_{1,2}$ relate the amplitudes of incoming and
outgoing waves $\exp(\pm ip_{Fx}x)$. So if the scattering
interface is shifted from $x=0$ to $x=x_{1,2}$ these matrices are
transformed as
\begin{gather}
\hat S_{1,x=x_1}= \hat\Lambda_1 \hat S_{1,x=0} \hat\Lambda_1,
\\
\hat\Lambda_1=\left(
\begin{array}{ccc} e^{ip_{Fx}x_1}& 0 &0 \\ 0& e^{ip_{Fx}x_1}&0\\ 0& 0 &
e^{-ip_{Fx}x_1}\end{array}\right)
\end{gather}
and
\begin{gather}
\hat S_{2,x=x_2}= \hat\Lambda_2 \hat S_{2,x=0} \hat\Lambda_2,
\\ 
\hat\Lambda_2=\left(
\begin{array}{ccc} e^{-ip_{Fx}x_2}& 0 &0 \\ 0& e^{-ip_{Fx}x_2}&0\\ 0& 0 &
e^{ip_{Fx}x_2}\end{array}\right).
\end{gather}

\subsection{Supercurrent}
\label{subsecsupercur}

Combining the solution of  Eqs. (\ref{ls1})-(\ref{ls4}) with Eqs.
(\ref{62})-(\ref{63}) we find
\begin{widetext}
\begin{equation}
I_+=ieT\sum_{\omega_n>0,k}\frac{e^{-i\chi}\left( \frac{A(\hat S_2,a^2)A^*(\hat
S_2,a^{-2})-B(\hat S_2,a^2)B(\hat S_2,a^{-2})}{\Phi(\hat S_2,a^2)\Phi(\hat
S_2,a^{-2})}\right)-e^{i\chi}\left( \frac{A(\hat S_1,a^2)A^*(\hat
S_1,a^{-2})-B(\hat S_1,a^2)B(\hat S_1,a^{-2})}{\Phi(\hat S_1,a^2)\Phi(\hat
S_1,a^{-2})}\right)}{ e^{-i\chi}\left( \frac{A(\hat S_2,a^2)A^*(\hat
S_2,a^{-2})-B(\hat S_2,a^2)B(\hat S_2,a^{-2})}{\Phi(\hat S_2,a^2)\Phi(\hat
S_2,a^{-2})}\right)+e^{i\chi}\left( \frac{A(\hat S_1,a^2)A^*(\hat
S_1,a^{-2})-B(\hat S_1,a^2)B(\hat S_1,a^{-2})}{\Phi(\hat S_1,a^2)\Phi(\hat
S_1,a^{-2})}\right)+\Pi}\label{j+}
\end{equation}
where
\begin{equation}
\Pi=\frac{q^4 B(\hat S_1,a^2)B(\hat S_2,a^{2})-A(\hat S_1,a^2)A(\hat
S_2,a^{2})}{a^2\Phi(\hat S_1,a^{2})\Phi(\hat S_2,a^{2})}+\frac{a^2\left[q^{-4}
B(\hat S_1,a^{-2})B(\hat S_2,a^{-2})-A^*(\hat S_1,a^{-2})A^*(\hat
S_2,a^{-2})\right]}{\Phi(\hat S_1,a^{-2})\Phi(\hat S_2,a^{-2})} \label{Pi}
\end{equation}
\end{widetext}
and
\begin{multline}
A(\hat S,a^{2})=-s_{33}+
\\a^2\left[
s_{31}(s_{13}s_{22}^*-s_{23}s_{21}^*)+s_{32}(s_{23}s_{11}^*-s_{13}s_{12}^*)
\right.
\\ 
\left. + s_{33}(|s_{12}|^2+|s_{21}|^2
-s_{11}s_{22}^*-s_{22}s_{11}^*) \right]+
\\
a^4\left[s_{12}^*s_{21}^*-s_{11}^*s_{22}^* \right]\left[
s_{31}(s_{23}s_{12}-s_{13}s_{22})+\right.
\\
\left.s_{32}(s_{13}s_{21}-s_{23}s_{11})+ s_{33}(s_{11}s_{22}-s_{12}s_{21})
\right],
\end{multline}
\begin{multline}
B(\hat S,a^2)=1+
\\
a^2\left(
s_{11}s_{22}^*+s_{22}s_{11}^*-|s_{12}|^2-|s_{21}|^2\right)
\\
+a^4\left(
s_{12}s_{21}-s_{11}s_{22}\right)\left(s_{12}^*s_{21}^*-s_{11}^*s_{22}^*\right),
\end{multline}
\begin{multline}
\Phi(\hat S,a^{2})=s_{13}^*s_{32}-s_{23}^*s_{31}+
\\
a^2\left[
s_{31}s_{13}^*(s_{22}s_{21}^*-s_{12}s_{22}^*) +
s_{32}s_{23}^*(s_{21}s_{11}^*-s_{11}s_{12}^*) \right.
\\
\left.
+s_{31}s_{23}^*(|s_{12}|^2-s_{22}s_{11}^*)  +
s_{32}s_{13}^*(s_{11}s_{22}^*-|s_{21}|^2)\right].
\end{multline}
Making use of the parameterization (\ref{shalf2}) of the scattering matrices we
eventually arrive at the general expression for the Josephson current
\begin{equation}
I=-\frac{8eT}{\Delta^2}\sum_{\omega_n>0,k}\frac{ \omega_n^2 \sin\tilde\chi}{
W-(4\omega_n^2\cos\tilde \chi/\Delta^2) }, \label{jchm}
\end{equation}
where
\begin{multline}
W=\frac{1}{2 {\cal D}_{12}}\left\{ q^4
\left[(a^{-1}-a)(a^{-1}-a\cos^2\nu_1)+|\tilde \alpha_1|^2
\right]\times\right.
\\
\times\left[(a^{-1}-a)(a^{-1}-a\cos^2\nu_2)+|\tilde \alpha_2|^2 \right]+
\label{wex}
\\ 
q^{-4} \left[(a^{-1}-a)(a^{-1}\cos^2\nu_1-a)+|\tilde
\alpha_1|^2 \right]\times
\\
\times\left[(a^{-1}-a)(a^{-1}\cos^2\nu_2-a)+|\tilde \alpha_2|^2
\right]-
\\
e^{i\varphi}\left((a^{-1}-a)^2 \cos\nu_1 +\tilde \alpha_1^2\right)
\left((a^{-1}-a)^2 \cos\nu_2 +\tilde \alpha_2^2\right)-
\\
e^{-i\varphi}\left((a^{-1}-a)^2 \cos\nu_1 +\tilde \alpha_1^{*2}\right)
\left. \left((a^{-1}-a)^2 \cos\nu_2 +\tilde
\alpha_2^{*2}\right)\right\}.
\end{multline}
Here $\varphi=-2\zeta_1-2\zeta_2+2p_{Fx}d$ is the quantum mechanical phase
corresponding to electron making a cycle between the two interfaces. As we
already discussed, in the many channel limit it is appropriate to average the
result (\ref{jchm}), (\ref{wex}) over quickly oscillating phase $\varphi$. This
averaging is accomplished with the aid of the relationship
\begin{equation}\int\limits_0^{2\pi}
\frac{1}{A+Be^{i\varphi}+C
e^{-i\varphi}}\frac{d\varphi}{2\pi}=\frac{1}{\sqrt{A^2-4BC}}. \label{ap}
\end{equation}
It is satisfactory to observe that after such averaging in Eqs. (\ref{jchm}),
(\ref{wex}) the general expression for the Josephson current exactly coincides
with Eqs. \eqref{jshs}-\eqref{betazeta} derived in Sec. \ref{secquas} within
our quasiclassical analysis.

\subsection{Weak tunneling limit}

Let us first analyze the above general results in
the limit of low interface transmissions $\nu_1,\nu_2 \ll 1$.
In this case Eq. (\ref{wex}) reduces to
\begin{multline}
W=\frac{1}{{\cal D}_{12}\Delta^4} \left\{ 16\left(\cosh\frac{2\omega_n
d}{v_{Fx}}-\cos\varphi \right)\times \right.
\\
\times
\left(\omega_n^2+\Delta^2(\Rea\alpha_1)^2\right)\left(
\omega_n^2+\Delta^2(\Rea\alpha_2)^2\right)  +
\\ 
2\omega_n^2\left[ \nu_1^4\left(\omega_n^2+\Delta^2(\Rea\alpha_2)^2\right)
+ \nu_2^4\left(\omega_n^2+\Delta^2(\Rea\alpha_1)^2\right)\right.
\\
\left.+2\nu_1^2\nu_2^2
(\omega_n^2+ \Delta^2)\right]
\\
+2\nu_1^4 \Delta^2(\Img\alpha_1)^2
\left(\omega_n^2+\Delta^2(\Rea\alpha_2)^2\right)
\\
+2\nu_2^4 \Delta^2(\Img\alpha_2)^2
\left(\omega_n^2+\Delta^2(\Rea\alpha_1)^2\right)
\\
+
4\nu_1^2\nu_2^2\Delta^4\Img\alpha_1\Rea\alpha_1\Img\alpha_2\Rea\alpha_2
\\
- 8\Delta^2\omega_n^2\left(\Img\alpha_1\Rea\alpha_1\nu_1^2+
 \Img\alpha_2\Rea\alpha_2\nu_2^2\right)\sin\varphi\bigg\}.
\end{multline}

\begin{figure}
\includegraphics{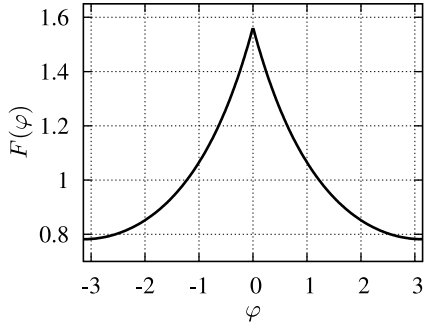}
\caption{The function $F(\varphi)$ defined in Eq. (\ref{fphi}).}
\label{figfphi}
\end{figure}

The resulting expression for the current for long $d\gg\xi_0$ junctions is
given by
\begin{widetext}
\begin{equation}
I=-\frac{eT}{2}\sum_{\omega_n>0,k} \dfrac{\Delta^2 \omega_n^2
{\cal D}_{12}\sin\tilde\chi}{
[\omega_n^2+\Delta^2(\Rea\alpha_1)^2][\omega_n^2+\Delta^2(\Rea\alpha_2)^2]
\left(\cosh\frac{2\omega_n d}{v_{Fx}}-\cos\varphi_k \right)} .
\end{equation}
\end{widetext}
Here and below we explicitly indicate the dependence of the
scattering phase $\varphi_k$ on the channel number $k$. At $T=0$
we obtain
\begin{equation}
I=-\dfrac{7\zeta(3) e }{4\pi \Delta^2 d^3} \sum_{k}
 \dfrac{v_{Fx }^3 {\cal D}_{12} \sin\tilde\chi}{
|\tilde\alpha_1|^2|\tilde\alpha_2|^2}F(\varphi_k).\label{d3f}
\end{equation}
Comparing this exact result with its quasiclassical analogue
(\ref{d3}) we observe that in Eq. (\ref{d3f}) the contributions of
different channels are weighted by the function $F(\varphi_k)$
which reads (see also Fig. \ref{figfphi})
\begin{equation}
F(\varphi)=\frac{2\varphi}{21\zeta(3)\sin\varphi}(\pi-|\varphi|)
(2\pi-|\varphi|), \label{fphi}
\end{equation}
where $-\pi\le \varphi\le \pi$. The average of this function over
this phase interval equals to unity.
This relatively weak modulation of the Josephson current in our
case is in a drastic contrast with the pronounced resonant
behavior of the zero-temperature current in conventional SNS
junctions, see, e.g., Ref. \onlinecite{GZ}. This difference is
formally due to the presence of $\omega_n^2$ in the expression for
the Josephson current (\ref{jchm}).

Note that although in the limit $d \gg \xi_0$ resonant effects
remain insignificant they gain importance for shorter junctions $d
\lesssim\xi_0$ which will be considered below. As before, we stick
to the case of weakly transmitting boundaries (small $\nu,\beta$)
assuming for simplicity $\Rea\alpha_1=\Rea\alpha_2=1$. In the
many-channel limit we have for $\xi_0 (\nu_1^2+\nu_2^2)\ll d
\lesssim \xi_0$
\begin{multline}
I=\frac{e\Delta}{32\pi^2 T d}{\rm Im}\,\psi'\left(\frac{1+i(\Delta/\pi T)
}{2}\right)\times\label{pgf}
\\
\times \sum_k \nu_1^2\nu_2^2 |\beta_1|
|\beta_2| v_{Fx} \sin\tilde \chi.
\end{multline}
Here $\psi'(z)=d^2 \ln \Gamma(z)/dz^2$ is the polygamma-function. In the limit
 $T\ll\Delta$ Eq. (\ref{pgf}) yields
\begin{equation}
I=-\frac{e}{16 \pi d}\sum_k \nu_1^2\nu_2^2 |\beta_1| |\beta_2| v_{Fx}
\sin\tilde \chi.
\end{equation}
We observe that at $T \to 0$ the Josephson current grows with decreasing
$d$ as $I \propto 1/d$. This dependence persists down to
$d\sim \xi_0 (\nu_1^2+\nu_2^2)$. At even smaller $d \ll
\xi_0(\nu_1^2+\nu_2^2)$
we obtain
\begin{equation}
I=-eT\Delta^2\sum_{\omega_n>0,k}\frac{\omega_n \nu_1^2\nu_2^2 |\beta_1|
|\beta_2|\sin\tilde \chi}{(\omega_n^2+\Delta^2)^{3/2}(\nu_1^2+\nu_2^2)},
\label{mch2}
\end{equation}
which yields at $T\ll\Delta$
\begin{equation}
I=-\frac{e\Delta}{2\pi}\sum_{k}\frac{\nu_1^2\nu_2^2 |\beta_1|
|\beta_2|\sin\tilde \chi}{\nu_1^2+\nu_2^2}.
\end{equation}

The above results correspond to effective averaging of the
$\varphi$-dependent Josephson current. Let us now see how these
results get modified if the resonance condition $\cos\varphi=1$
holds for some of the conducting channels. In this case we have
\begin{equation}
I=-\frac{e}{32d^2\Delta}\frac{\sinh\frac{\Delta}{T}- \frac{\Delta}{T}}{1+ \cosh
\frac{\Delta}{T}} {\sum_k}^{\prime} \nu_1^2\nu_2^2 |\beta_1| |\beta_2| v_{Fx}^2
\sin\tilde \chi ,\label{res1}
\end{equation}
for $d \gg \xi_0 (\nu_1^2+\nu_2^2)$ and
\begin{equation}
I=-e\Delta \tanh\frac{\Delta}{2T}{\sum_k}^{\prime} \frac{\nu_1^2\nu_2^2
|\beta_1| |\beta_2|\sin\tilde \chi}{(\nu_1^2+\nu_2^2)^2}.\label{res2}
\end{equation}
in the limit $d \ll \xi_0 (\nu_1^2+\nu_2^2)$. Here the sum ${\sum}_k^{\prime}$
runs over the resonant channels only.

The resonant currents (\ref{res1}) and (\ref{res2}) turn out to be
much larger than the corresponding phase-averaged contributions
(\ref{pgf}), (\ref{mch2}). Correspondingly, the resonances are
quite narrow: $\delta\varphi\sim d/\xi_0$ and $\delta\varphi\sim
\nu_1^2+\nu_2^2$ for longer and shorter junctions. The
off-resonant currents are smaller in the measure of $d^2/\xi_0^2$
and $(\nu_1^2+\nu_2^2)^2$ respectively.

\subsection{High transmission limit}
For completeness let us analyze the case of fully transmitting interfaces, i.e.
$\sin^2\nu_1=\sin^2\nu_2=1$. For short $d\ll \xi_0$ junction the Josephson
current is defined by Eq. (\ref{jchm}) with
\begin{multline}
W=\frac{2}{|\beta_1||\beta_2|}\left(
|\alpha_1|^2|\alpha_2|^2\sin^2\frac{\varphi}{2}+4\frac{\omega_n^4}{\Delta^4}
\right.\label{tr1}
\\ 
\left. +\left(2+|\alpha_1|^2+|\alpha_2|^2\right)
\frac{\omega_n^2}{\Delta^2}\right)
\end{multline}
Here the initial phase $\varphi$ is shifted by $2\arg\alpha_{1}+
2\arg\alpha_{2}$. In particular, at low temperatures $T\ll \Delta$ and for
small $\beta_{1,2}$ we have
\begin{equation}
I=-\frac{e\Delta}{4}\sum_k\frac{|\beta_1||\beta_2|\sin\tilde
\chi}{\sqrt{1+\left|\sin\frac{\varphi_k}{2} \right|}}.
\end{equation}

In the opposite long junction limit  $d\gg \xi_0$ and at $T\ll
\Delta$ the Josephson current is again given by Eq. (\ref{jchm}) where one
should substitute
\begin{multline}
W=\frac{1}{|\beta_1||\beta_2|} \left[\left(\cosh\frac{ 2\omega_n
d}{v_{Fx}}-\cos\varphi\right)|\alpha_1|^2|\alpha_2|^2 \right.
\\
+\frac{2\omega_n^2}{\Delta^2}(2+|\alpha_1^2|+|\alpha_2|^2)\cosh\frac{ 2\omega_n
d}{v_{Fx}}
\\
\left.+2\frac{\omega_n}{\Delta} \sinh\frac{ 2\omega_n
d}{v_{Fx}} (|\alpha_1|^2+|\alpha_2|^2)\right] \label{tr2}
\end{multline}
For $|\alpha_{1,2}|\sim 1$ we again recover Eq.(\ref{d3f}), in which we should
replace $\tilde\alpha_{1,2}$ by $\alpha_{1,2}$.

Of a special interest is the case of identical interfaces  with
$|\beta_{1,2}|=1$ which corresponds to equal probabilities for spin-up and
spin-down electrons to penetrate into the half-metal. In this case
Eqs.(\ref{tr1}), (\ref{tr2}) yield
\begin{equation}
I=-\frac{e\Delta}{2} \mathcal{N} {\rm sgn}
\chi\cos\frac{\chi}{2} \label{KO}
\end{equation}
for $d \ll \xi_0$ and
\begin{equation}
I=\frac{e}{2\pi d}   (\chi -\pi  \sgn \chi ) \sum\limits_k v_{Fx}
\label{IK}
\end{equation}
for $d \gg \xi_0$. Here the Josephson phase difference is restricted within
the interval $-\pi\le \chi\le \pi$ and the total number of conducting
channels in our SHS junction is denoted by ${\cal N}$. One easily recognizes 
that Eqs. (\ref{KO}) and (\ref{IK}) are
nothing but the $\pi$-shifted Kulik-Omelyanchuk \cite{KO} and Ishii-Kulik
\cite{IK} current-phase relations respectively for short and long SNS
junctions. It is also interesting to observe that for $\mathcal{N}=2$ (i.e. in the
single channel limit with two spin directions) Eq. (\ref{IK})
coincides with the result for the Josephson current in long SNS junctions
embedded in a superconducting ring with odd number of electrons \cite{SZ}. It
also follows from Eq. (\ref{tr2}) that the current-phase relation (\ref{IK})
for long junctions $d\gg \xi_0$ sets in only in the narrow parameter region
$|\alpha_1|^2|\alpha_2|^2\lesssim \xi_0^2/d^2$. Correspondingly, the
dependence of the Josephson current on the H-layer thickness changes from $I
\propto 1/d^3$ to $I \propto 1/d$.

\subsection{Andreev states}
The expressions for the Green functions and for the
Josephson current derived above enable
us to obtain the spectrum of subgap Andreev bound states inside
the half-metal. Substituting $\omega_n\rightarrow -iE$ we easily
find the poles of the Green function which provide the required
spectrum. For junctions $d\gg \xi_0$ with low transparency (small
$\beta,\nu$) it is governed by the equation
\begin{equation}
\cos\left(\frac{2Ed}{v_{Fx}}\right)-\cos\varphi+ \frac{E^2 {\cal D}_{12}
\cos\tilde\chi}{4\Delta^2 ({\rm Re}\,\alpha_1)^2 ({\rm
Re}\,\alpha_2)^2}=0.\label{and1}
\end{equation}
Here we assume that $E\ll \Delta$ and ${\rm Re}\,\alpha_1\sim {\rm
Re}\,\alpha_2\sim 1$. Though the third term in this equation is
much smaller than the other two, it is important since it
determines the dependence of the spectrum on the Josephson phase
$\chi$. Note that the above equation remains valid for all values
of $\varphi$ except for an immediate vicinity of the resonance
$\cos\varphi=1$ where the full expression for $W$ has to be taken
into account. In the latter limit the result becomes rather
cumbersome and is omitted here.

It is instructive to compare the above expression with that for
Andreev bound states in conventional SNS-junctions in the limit of
low transmissions $D_{1,2}$ of NS interfaces, see, e.g., Ref.
\onlinecite{GZ}. For long SNS junctions $d\gg \xi_0$ we have
\begin{equation}
\cos\left(\frac{2Ed}{v_{Fx}}\right)-\cos\varphi+\frac{ D_1
D_2}{4}\cos\chi=0. \label{andn1}
\end{equation}
As before, the parameter $\varphi$ is defined as
$\varphi=\varphi_0+2p_{Fx}d$. In the limit $D_{1,2}=0$ this
equation describes particle- and hole-like excitations inside the
normal layer with impenetrable boundaries. At small but non-zero
$D_{1,2}$ these states get modified due to the superconducting
proximity effect. We observe that in the presence of triplet pairing
states inside our SHS junction Eq. (\ref{and1}) has essentially the
same structure as Eq. (\ref{andn1}) describing singlet pairing
states except for an additional small factor $E^2/\Delta^2$
entering the $\chi$-dependent term in the triplet case.

In the opposite limit of short junctions $d\ll \xi_0$ the spectrum
of Andreev levels can be found analogously. For SHS junctions we
obtain
\begin{multline}
(1-\cos\varphi)\left( ({\rm Re}\,\alpha_1)^2
-\frac{E^2}{\Delta^2}\right) \left( ({\rm Re}\,\alpha_2)^2
-\frac{E^2}{\Delta^2}\right)
+\\+
\frac{E^2}{4\Delta^2} {\cal D}_{12} \cos\tilde\chi=0. \label{and2}
\end{multline}
In the case  ${\rm Re}\,\alpha_{1,2}\approx 1$ this result
describes excitations that are slightly below the gap.

For comparison we also recall the well known expression which defines the
spectrum of Andreev levels for
short ($d\ll \xi_0$) conventional SNS junctions:
\begin{multline}
\cos\varphi\left(\frac{E^2}{\Delta^2}-1\right)\left(
1-\frac{(D_1+D_2)^2}{8}\right) 
+\\ +
1-\frac{E^2}{\Delta^2}
-\frac{D_1 D_2}{4}(1-\cos\chi )=0.
\end{multline}
Similarly to  Eq. (\ref{and2}), this equation describes the states with
energies slightly below the gap. We would also like to point out that Andreev
states in SNS junctions are doubly degenerate. This degeneracy is in general
lifted in the case of triplet pairing states in SHS junctions, see also
\cite{E04}.

\begin{figure}
\includegraphics{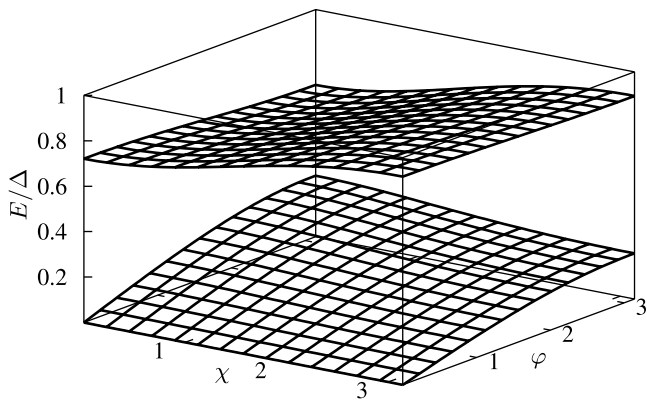}
\caption{The higher and lower Andreev levels in short SHS
junctions with highly transparent interfaces for
$|\beta_1|=|\beta_2|=0.8$} \label{fand1}
\end{figure}

\begin{figure}
\includegraphics{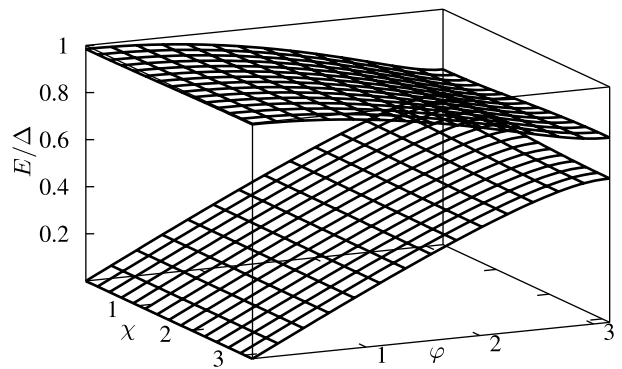}
\caption{The same as in Fig. \ref{fand1} for
$|\beta_1|=|\beta_2|=0.2$} \label{fand2}
\end{figure}

Turning to the limit of highly transparent interfaces, in the
limit of long ($d\gg \xi_0$) SHS junctions for $|\alpha_{1,2}|\sim
1$ we again arrive at Eq.(\ref{and1}) where one should only
replace $({\rm Re}\,\alpha_{1,2})^2$ by $|\alpha_{1,2}|^2/4$. In
the opposite short junction limit $d\ll \xi_0$ we obtain the
following equation, which is valid for arbitrary relationship
between parameters $\alpha$ and $\beta$
\begin{multline}
4\frac{E^4}{\Delta^4}-\left(2-2|\beta_1||\beta_2|\cos\tilde\chi+|\alpha_1|^2+
|\alpha_2|^2\right)\frac{E^2}{\Delta^2}
+\\+
|\alpha_1|^2|\alpha_2|^2\sin^2\frac{\varphi}{2}=0.
\end{multline}
For positive energies $0\le E\le
\Delta$ it has two solutions which are depicted in Figs.
\ref{fand1} and \ref{fand2} for the case of fully transparent SH
interfaces $\sin^2\nu_{1,2}=1$.

We observe that for $|\beta_{1,2}|\approx 1$ (see Fig.
\ref{fand1}) the position of Andreev levels significantly depends
on the Josephson phase $\tilde\chi$ whereas the
$\varphi$-dependence remains not very pronounced. In the limit
$|\beta_{1,2}|=1$ the energy of the lower Andreev level reduces to
zero, while the energy for the upper one is determined by a simple
formula
\begin{equation}
E=\Delta|\sin(\tilde\chi/2)|.
\end{equation}
Note that this dependence is just the $\pi$-shifted one as
compared to the well known dependence $E=\Delta|\cos(\chi/2)|$ for
the doubly degenerate Andreev level in short SNS junctions.

In the limit of small spin-flip factors $|\beta_{1,2}|$ the
behavior of Andreev levels changes considerably, as illustrated in
Fig. \ref{fand2}. In this case the dependence of the level
positions on the Josephson phase $\tilde\chi$ is much weaker,
whereas their $\varphi$-dependence becomes more significant.

\begin{figure}
\includegraphics{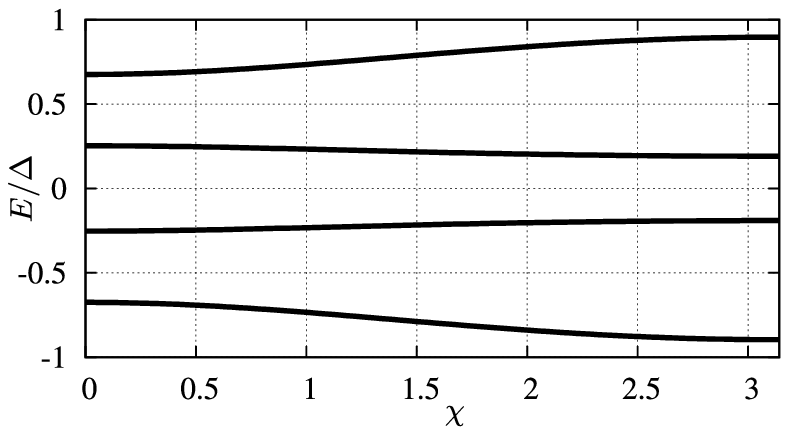}
\includegraphics{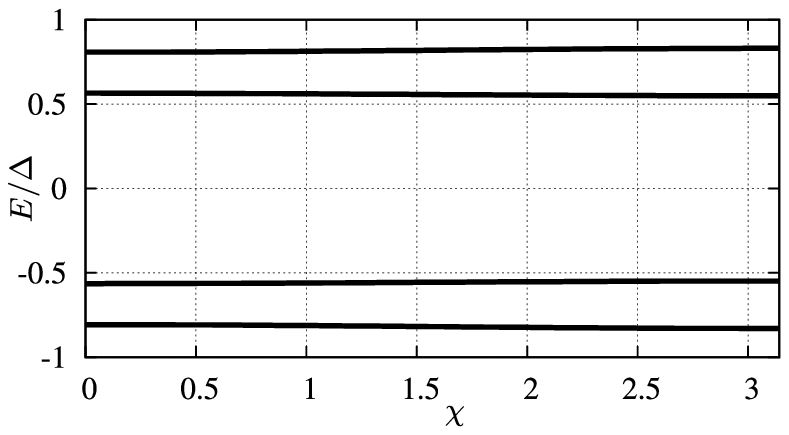}
\caption{The structure of Andreev bound states in short SHS
junctions with highly transparent interfaces for
$|\beta_1|=|\beta_2|=0.8$ (upper figure) and
$|\beta_1|=|\beta_2|=0.2$ (lower figure) at $\varphi=2.5$ }
\label{fand3}
\end{figure}

Clearly, Andreev levels with negative energies $-\Delta \le E\le
0$ are fully symmetric with respect to the Fermi level and, hence,
show exactly the same features, as illustrated by Fig.
\ref{fand3}. We again observe that the $\tilde\chi$-dependence of
all four Andreev states is rather pronounced for large values of
the spin-flip parameter $|\beta|$ and it almost disappears for
small values of $|\beta|$.

\section{Summary}
\label{secsumm}
In this paper we have developed a detailed
microscopic theory describing dc Josephson effect and Andreev bound states
in superconducting junctions with a half metal. The possibility to
pass supercurrent through such SHS junctions is provided by
spin-flip scattering at both SH interfaces. As a result,
superconducting correlations penetrate into the half-metal in the
form of triplet pairing states which can carry non-vanishing
supercurrent across the system. Our general results for the
Josephson current, Eqs. (\ref{jshs}) and (\ref{jchm}), provide
a detailed description of the effect for sufficiently clean metals
and at arbitrary transmissions $\sin^2 \nu_{1,2}$ of both
interfaces as well as at arbitrary values of the spin-flip
parameters $\beta_{1,2}$. In the tunneling limit our results
reduce to those of Ref. \onlinecite{Eschrig07} derived
perturbatively in the interface transmissions. Our approach allows
to go beyond the perturbation theory in $\nu_{1,2}$ and to fully
account for all orders in interface transmissions. This
non-perturbative analysis is unavoidable not only at high
transmissions but also in the tunneling limit in the case of
sufficiently short junctions in order to eliminate an intrinsic
divergence of perturbative results at small $d$. At
$T=0$ the Josephson current depends on the H-layer thickness as
$I \propto 1/d^3$ for $d\gg\xi_0$. This dependence then crosses over to
$I \propto 1/d$ for $\xi_0 (\nu_1^2+\nu_2^2)\ll d \ll \xi_0$ (see Eq.
(\ref{pgf})). Finally, the current becomes $d$-independent at $d\ll \xi_0
(\nu_1^2+\nu_2^2)$ (see Eq.(\ref{mch2})). This saturation of the
$d$-dependence of the Josephson current is
accompanied by the change of its scaling with $\nu$ from
 $I \propto \nu^4$ to $I \propto \nu^2$. For short
junctions $d\ll\xi_0$ with few conducting channels resonant
effects play an important role
causing substantial enhancement of the Josephson current, see
Eqs.(\ref{res1}) and (\ref{res2}).

Our analysis demonstrates that the behavior of both
the Josephson current and Andreev bound states in SHS structures
crucially depends on the spin-flip parameters $\beta_{1,2}$.
Similarly to Refs. \onlinecite{Eschrig03,Eschrig07} we observe
that for incomplete spin-flip scattering the temperature dependence
of the critical Josephson current is non-monotonous with a maximum
at non-zero $T$, see Fig. 3. However, this feature disappears
completely for $\beta_{1,2} \to 1$ and the Josephson current monotonously
increases with decreasing $T$ in this limit. Another striking feature
of our results is that SHS junctions are characterized by the
the sinusoidal current-phase relation at any temperature and interface
transmissions provided the spin-flip parameters $\beta_{1,2}$ are not
very close to unity. Substantial deviations of the current-phase relation
occur only provided (i) interface transmissions remain high, (ii)
temperature remains low and (iii) the spin-flip parameters
$\beta_{1,2}$ are very close to one, see Fig. 2.

It also follows from our analysis that for symmetric SHS junctions
(meaning that electron scattering at both interfaces is described by identical
$S$-matrices) the $\pi$-junction state is usually realized. However,
in a general case the Josephson current in SHS junctions does not necessarily
show the $\pi$-junction behavior, the current-phase relation is
characterized by an arbitrary phase shift, see Eq.
(\ref{betazeta}). It is also interesting to observe that
in the case of fully transmitting interfaces, at $T=0$ and
for $|\beta_{1,2}|=1$
our results reduce to very simple dependencies, Eqs. (\ref{KO}) and (\ref{IK}),
which are essentially the $\pi$-shifted Kulik-Omelyanchuk \cite{KO}
and Ishii-Kulik \cite{IK} current-phase relations respectively for
short and long SNS junctions.

We believe that our predictions can help to identify triplet pairing
current states in future experiments with SHS structures.

\centerline{\bf Acknowledgments}

\vspace{0.5cm}

We acknowledge useful discussions with M. Eschrig. This work was
supported in part by RFBR grant 06-02-17459.

\end{document}